\documentclass[aip,reprint,superscriptaddress]{revtex4-1}  
\usepackage{graphicx}  
\usepackage{dcolumn}   
\usepackage{bm}        
\usepackage{amssymb}   
\usepackage{epsfig,amsmath,amssymb,color,float,setspace}
\usepackage[utf8]{inputenc}
\usepackage[T1]{fontenc}
\usepackage{titlesec}
\usepackage{booktabs,verbatim}
\usepackage{graphicx}
\usepackage{epstopdf}
\usepackage[scriptsize]{caption}
\newcommand{\squeezeup}{\vspace{-2.5mm}}

\begin{document}
\title{Magnetic properties of low-moment ferrimagnetic Heusler Cr$_2$CoGa thin films grown by molecular beam epitaxy}
\author{Michelle E. Jamer}
\affiliation{NIST Center for Neutron Research, National Institute of Standards and Technology, Gaithersburg, MD 20899, USA}
\affiliation{Department of Physics, Northeastern University, Boston, MA 02115, USA}
\author{George E. Sterbinsky}
\affiliation{Advanced Photon Source, Argonne National Laboratory, Argonne IL, 60439, USA}
\author{Gregory M. Stephen}
\affiliation{Department of Physics, Northeastern University, Boston, MA 02115, USA}
\author{Matthew C. DeCapua}
\affiliation{Department of Physics, Northeastern University, Boston, MA 02115, USA}
\author{Gabriel Player}
\affiliation{Department of Physics, Northeastern University, Boston, MA 02115, USA}
\author{Don Heiman}
\affiliation{Department of Physics, Northeastern University, Boston, MA 02115, USA}
\date{\today}
\begin{abstract}
Recently, theorists have predicted many materials with a low magnetic moment and large spin-polarization for spintronic applications. These compounds are predicted to form in the inverse Heusler structure, however, many of these compounds have been found to phase segregate. In this study, ordered Cr$_2$CoGa thin films were synthesized without phase segregation using molecular beam epitaxy. The present as-grown films exhibit a low magnetic moment from antiferromagnetically coupled Cr and Co atoms as measured with SQUID magnetometry and soft X-ray magnetic circular dichroism. Electrical measurements demonstrated a thermally-activated semiconductor-like resistivity component with an activation energy of 87 meV. These results confirm spin gapless semiconducting behavior, which makes these thin films well positioned for future devices.
\end{abstract}

\maketitle
Spin gapless semiconductors (SGS) have been predicted to merge the properties of gapless semiconductors and half-metallic magnets, which would be beneficial in various magnetoelectronic applications.\cite{Felser1,Wang2008} The schematic of the density of states (DOS) is seen in Fig. 1(a), where the spin up (orange) band is the majority state and the spin down (green) band is the minority state, and the Fermi energy (E$_F$) is placed so that the majority band acts as a gapless semiconductor.\cite{Wang1,Wang2} There has been experimental and theoretical research on SGS materials that are either ferromagnetic or ferrimagnetic with a large overall magnetic moment.\cite{Mn2CoAl1,Jamer1,SKAFT,Jamer3,Jamer5} However, for high density spintronic technologies, a low magnetic moment is preferable in order to minimize interactions while maintaining spin polarization as in the case of Mn$_2$Ru$_x$Ga.\cite{MnRuGa} Binary V$_3$Al with a Heusler structure has been synthesized in bulk and has been shown to be a traditional G-type N\a'eel antiferromagnet with two identical compensating vanadium moments, and therefore does not have the ability to be spin polarized due to its symmetrical density of states.\cite{Jamer4, CoeyBook}
\begin{figure}
\begin{centering}
\includegraphics[width=0.5\textwidth]{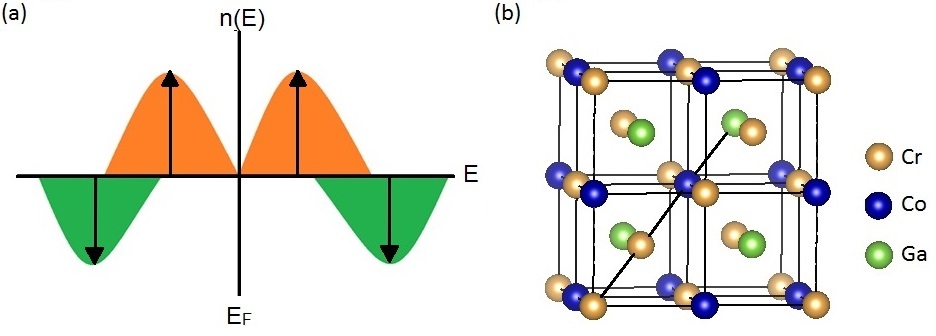}
\vspace*{-5mm}
\setlength{\belowcaptionskip}{-16pt}
\caption{Illustrations of (a) the SGS density of states showing the majority and minority bands, and (b) the inverse Heusler XA F\a=43m lattice of Cr$_2$CoGa.}
\end{centering}
\end{figure}

Cr$_2$CoGa has been predicted to be an inverse Heusler half-metallic compound with low net moment equal to 0.09 $\mu_B$ per formula unit (f.u.) but large Curie temperature between 1300-1600 K.\cite{SKAFT, SkaftAPL,Galanakisa, Meinertb, Meinertc,Deka,CoeyAPL,Graf} The inverse Heusler (XA) phase has a space group F\a=43m, which varies slightly from the full Heusler (L2$_1$) phase with space group Fm\a=3m, seen in Fig. 1(b).\cite{Bansil1999} The difference between the two structures leads to an obvious change in the body diagonal along the <111> direction, indicated by a bold line in Fig. 1(b). The L2$_1$ structure has an atomic configuration of X-Y-X-Z along the <111> diagonal, whereas the XA structure has X-X-Y-Z.\cite{SKAFT,Mn2CoAl1} The activation energy for the formation of Cr$_2$CoGa in the XA lattice is lower than the L2$_1$ structure, however, the activation energy is still positive (+0.08 eV/f.u. and +0.7 eV/f.u., respectively), indicating that phase segregation will occur for equilibrium conditions.\cite{Meinertb, Meinertc,Galanakisa} However, Cr$_2$CoGa with phase segregates was previously synthesized by Feng et al. using molecular beam epitaxy (MBE) at 450 $^o$C.\cite{Feng, Meinertc} In the present study, singular phase inverse Heusler Cr$_2$CoGa thin films were achieved via MBE growth without phase segregates present in the unannealed sample. The nonequilibrium growth technique of MBE is able to lower the energy of formation of the Cr$_2$CoGa, which prevents the phase from segregating.\cite{Graf} The resulting film was highly ordered and possessed a low magnetic moment and large Curie temperature (T$_C$). The measurements indicate that the low moment ferrimagnetic state of single phase Cr$_2$CoGa persists well above room temperature making it attractive for applications.

X-ray diffraction (XRD) was used to investigate the crystallographic phases using Cu-K$\alpha$ radiation. The magnetic properties were measured using a magnetic properties measurement system superconducting quantum interface device (SQUID) magnetometer in fields up to 5 T and between 2-400 K.  In addition, the atomic moments were measured using X-ray magnetic circular dichroism (XMCD) at beamline U4B at the National Synchrotron Light Source (NSLS). The XMCD measurements were taken at 70$\%$ circular polarization using total electron yield mode with an applied field of 1.5 T at 10 and 300 K. Element specific resolution was achieved by tuning the energy to the Cr (565-600 eV) and Co (760-820 eV) L$_{2,3}$ edges. Electrical measurements in the range T = 2 - 300 K and $\mu_0$H = 5 T were made using a modified transport probe designed for use in the SQUID magnetometer.\cite{Assaf}
\begin{figure}
\begin{centering}
\includegraphics[width=0.4\textwidth]{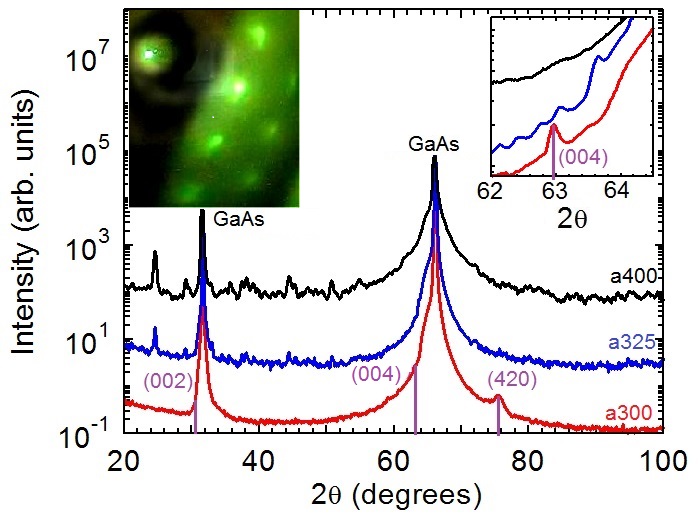}
\vspace*{-5mm}
\setlength{\belowcaptionskip}{-16pt}
\caption{X-ray diffractograms for as-grown and annealed samples of Cr$_2$CoGa films. Peaks are observed for the (002), (004) and (420) diffractions for the as-grown sample (a300). The left inset shows the RHEED pattern in the as-grown (a300) sample. The right inset shows the (004) peak near the GaAs substrate peak for the a300 sample plotted on a linear scale, which is at ~63.11$^o$. As the structure is annealed, more peaks appear in the diffractograms.}
\end{centering}
\end{figure}
Films of Cr$_2$CoGa ($\approx$ 52 nm) were grown on GaAs (001) substrates in an ultra-high-vacuum MBE apparatus using separate Knudsen effusion cells. The surface oxide of the GaAs substrates was removed.\cite{Jamer1} Reflection high energy electron diffraction (RHEED) patterns indicate ordering and alignment of the surface atoms. Stoichiometry was confirmed with energy dispersive spectroscopy. The thin films were further annealed post deposit to 325 $^o$C (a325) and 400 $^o$C (a400) for 30 minutes in ultra-high vacuum $\approx$ 10$^{-7}$ Pa to investigate the effects of thermal processing.

The XRD diffractograms of Cr$_2$CoGa are seen in Fig. 2, where the as-grown sample (a300) shows (002), (004), and (420) peaks that are consistent with the inverse Heusler lattice marked with purple lines. The left inset of Fig. 2 shows the RHEED pattern of the as-grown sample after growth, which shows partial epitaxial alignment on the (001) plane, indicating ordered polycrystallinity causing unique texturing of the sample.\cite{Jamer1} This strong texturing can account for lack of an observable (220) peak which is normally the largest. The calculated lattice constants from these XRD peaks indicates a tetragonal distortion with \textit{a} = 5.63 $\pm$ 0.11 \AA \hspace{1mm} and \textit{c} = 5.88 $\pm$ 0.15 \AA. The a325 XRD pattern suggests that phase segregation occurred with the emergence of new peaks, and the phase segregation becomes more apparent with the a400 sample. According to the diffractograms, the phase segregation from the Cr$_2$CoGa lattice increases as the annealing temperature increases. These phase segregates agree with previous results seen in Cr$_2$CoGa thin films and can be indexed to the Cr$_3$Ga+Co+Cr+GaCo phase segregates with the lower reaction energy (-0.38 eV/f.u.) than that of Cr$_2$CoGa.\cite{Feng, Meinertb}.

\begin{figure*}
\begin{centering}
\includegraphics[width=0.85\textwidth]{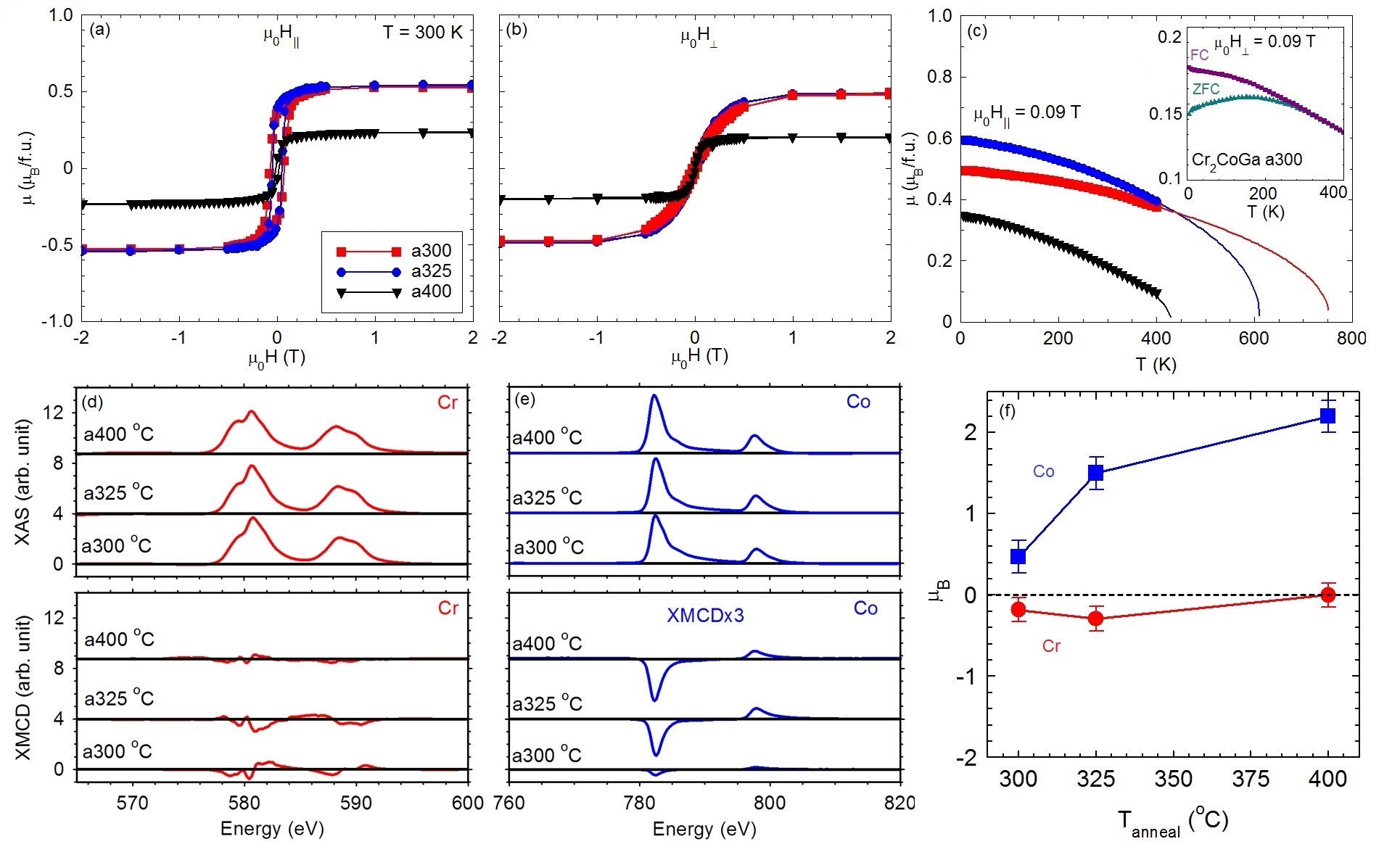}
\vspace*{-5mm}
\setlength{\belowcaptionskip}{-16pt}
\caption{The M(H) loops of Cr$_2$CoGa thin films annealed at 300-400 $^o$C measured in both the (a) parallel and (b) perpendicular field orientations relative to the film surface in Bohr magnetons per formula unit ($\mu_B$/f.u.) at room temperature T = 300 K. (c) The M(T) curves for Cr$_2$CoGa thin films for three annealing temperatures when the magnetic field is applied parallel to the films. The Curie transition temperatures are clearly above 400 K. The inset shows the zero-field cooled (ZFC) and field cooled (FC) measurements when the field was applied perpendicular to the a300 samples. All M(T) measurements were taken at $\mu_0$H = 0.09 T. The XAS and XMCD measurements for (d) Cr and (e) Co atoms in Cr$_2$CoGa as a function of annealing.(f) The extracted magnetic moments from the XAS and XMCD curves for both the Cr(red) and Co(blue) atoms as a function of annealing.}
\end{centering}
\end{figure*}
The magnetization ($\mu$) was measured as a function of magnetic field and is shown in Fig. 3 (a). As a function of annealing, the magnitude of the saturation moment first increases slightly from $\mu$ = 0.5 $\mu_B$/f.u. for annealing at 325 $^o$C, then decreases to $\mu$ = 0.2 $\mu_B$/f.u. for annealing at 400 $^o$C. The decrease in moment is attributed to a decrease in Cr$_2$CoGa fraction from phase segregation as indicated in the XRD. These figures also indicate a magnetic anisotropy as seen by the small difference in the approach to saturation for fields applied parallel and perpendicular relative to the film surface. At low fields the moment approaches saturation more quickly for the perpendicular orientation due to shape anisotropy. This anisotropy is expected and is a sign of texturing. The magnetization was also measured as a function of temperature, $\mu$(T), and field direction. Figure 3(c) displays $\mu$(T) between 5 and 400 K for the three samples at $\mu_0$H = 0.09 T, showing that the material has a large Curie temperature, T$_{C}$ greater than 400 K. The $\mu$(T) data was fit to the mean field model \begin{math}\mu(T) = \mu{_o}(1 - \frac{T}{T{_c}}){^{\gamma}}\end{math}, where $\gamma$ is a fitting parameter on the order of $\gamma$$\sim$ $\frac{1}{2}$. Table I lists results of the fits, where T$_C$ monotonically decreases from 750 to 440 K for increasing annealing temperature. Interestingly, for fields applied perpendicular to the film the $\mu$(T) data show differences for field cooling (FC) and zero field cooling (ZFC), where the ZFC data shows maxima in the 100 to 200 K range. This behavior can be attributed to nanoscale magnetic domain structure. The perpendicular magnetization data indicates that the thin films have some superparamagnetic behavior, with a blocking temperature of T$_B$ $\approx$ 150 K.

The total atomic moments of Cr and Co were measured through analysis of the XAS taken at the L$_3$ and L$_2$ edges with circularly polarized X-rays.\cite{Thole1992, Carra1993} In Fig. 3, the XAS measurements of the Cr (d upper) and Co (e upper) L$_3$ and L$_2$ edges are plotted as a function of energy. There are noticeable changes in the Cr XAS spectra as a function of annealing. The as-grown sample (a300) has a valence occupancy of Cr$^{3.1+}$, obtained by considering previous modeling of the Cr \textit{d}-orbitals \cite{Stavitski}. The valence then increases to Cr$^{3.4+}$ when annealed to 325 $^o$C, and finally to Cr$^{3.7+}$ when annealed to 400 $^o$C. Therefore, the Cr occupancies are changing as a function of annealing, indicating that there is phase segregation occurring at the higher annealing temperatures. The Co L-edges have a relatively constant valence Co$^{3+}$ and orbital configuration of (3d$^6$) as a function of annealing.\cite{Ravel2005}
\begin{figure}
\begin{centering}
\includegraphics[width=0.4\textwidth]{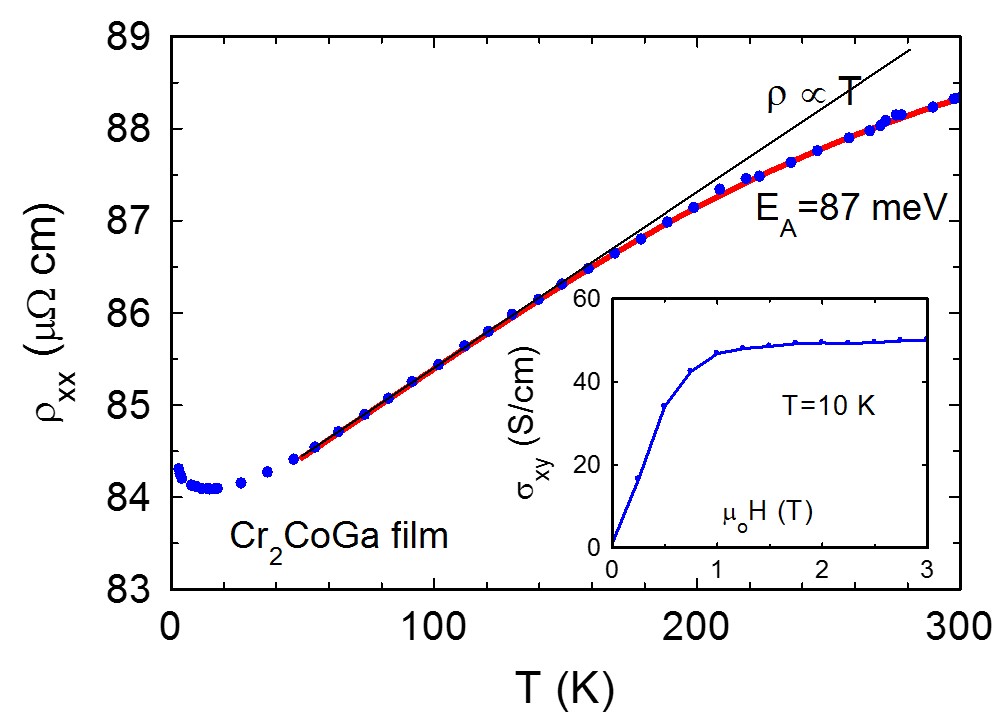}
\setlength{\belowcaptionskip}{-16pt}
\vspace*{-5mm}
\caption{Electrical transport of a 52 nm thick Cr$_2$CoGa film grown at 300 $^o$C, showing resistivity $\rho$ as a function of temperature from T = 2 to 300 K. The solid red curve fits the data using a thermal activation energy of 87 meV. ( inset) Magnetic field dependence of the Hall conductivity at T = 10 K displaying the anomalous Hall effect with a saturation conductivity $\sigma_{xy}$ = 50 S/cm.}
\end{centering}
\end{figure}
XMCD measurements were taken using opposite circular polarization and using positively and negatively applied fields in order to obtain XMCD spectra for the Cr (d lower) and Co edges (e lower), seen in Fig. 3. The Cr signal was quite low due to the antiferromagnetically coupled Cr atoms, and there was a branching ratio of L$_3$/L$_2$ = 1.53.\cite{LauBranching,Aksov2010,Wende} The extracted magnetic moments are plotted in Fig. 3(f), where the Co atomic moment increases by a factor of 7 between the  temperatures 300-325 $^o$C, further indicating phase segregates formed from the Heusler lattice after annealing at 325 $^o$C. The increasing magnetic moment supports the possible XRD phase segregations with an increasing Co phase concentration.\cite{Meinertc, Meinertb} Table I lists the magnetic moments measured using SQUID magnetometry and the XMCD-extracted Cr and Co magnetic moments as a function of annealing. The SQUID magnetic moments are quite low, between 0.21 and 0.46 $\mu_B$/f.u., while the magnetic moment of the Co atoms are quite large in comparison. These differences increase for increasing annealing temperatures. The difference in the moments resides primarily in the increasing moment on the Co atoms. These phase decompositions would have a large Co moment, however, the magnetic moment of the Cr would remain low due to its antiferromagnetic properties, which supports the low XMCD magnetic moment of Cr. Due to the structural disorder in the thin film, the magnetic moment is larger than the predicted 0.09 $\mu_B$/f.u. and has increased to 0.46 $\mu_B$/f.u.\cite{Deka}
\begin{table}
\begin{tabular}{@{}ccccc@{}}
\toprule
{T$_{\text{anneal}}$($^{\text{o}}$C}) & {M$_{\text{S}}$($\mu_B$/f.u.)} & {Cr($\mu_B$)} & {Co($\mu_B$)} & {T$_{\text{C}}$(K)} \\ \midrule
300     & 0.46    & -0.18    & 0.47     & 750    \\
325     & 0.57    & -0.29    & 1.5      & 640   \\
400     & 0.21    & -0       & 2.2      & 440  \\ \bottomrule
\end{tabular}
\caption{The magnetometry results of Cr$_2$CoGa as a function of annealing. Results from SQUID magnetometer measurements show low values of the magnetic moments. However, the XMCD results for the individual Co magnetic moment are quite large in comparison.} 
\label{tab:XI}
\squeezeup
\end{table}

The electrical properties of a Cr$_2$CoGa film were measured as a function of temperature. The room temperature resistivity was found to be $\rho_{300}$ $\approx$ 90 $\mu\Omega$-cm. Figure 4 shows that the temperature-dependent resistivity is metallic-like, but is composed of temperature-independent and temperature-dependent contributions $\rho$ = $\rho_0$ + $\rho$(T). The temperature dependent contribution $\rho$(T) is seen to be linear with T only over an intermediate range 50 - 200 K, where the carrier concentration is constant and mobility is affected by phonon scattering. It becomes sublinear above 200 K, due to an increase in the number of carriers. For this semiconductor-like activation of carriers where $\rho$(T) can be fit to
\begin{equation*}
\frac{-1}{\rho_{xx}}(T) = \sigma_{xx}(T) = n_m e\mu_{ph}(T)+n_a(T) e \mu_{a},
\end{equation*}
where $\mu_{ph}(T) = 1/(cT+d)$, cT is due to carrier-phonon scattering, and n$_m$ is the number of metallic carriers which remains constant. The fit assumed $\mu_{ph}(T)$ = $\mu_a$.\cite{Jamer1}  The number of thermally activated carriers in the second term varies as $n_a$(T) =  $exp(-\Delta/T)$ with activation temperature $\Delta$. The activation energy of the carriers was found to be 87 meV, which is on the same order as SGS Mn$_2$CoAl\cite{Jamer1,Mn2CoAlXu} and Ti$_2$MnAl.\cite{FengTi} The anomalous Hall effect (AHE) behavior shown in the inset of Fig. 4 reflects the measured magnetization seen in Fig. 3(b). The AHE conductivity was found to be $\sigma_{xy}$ = 50 S/cm, a small value consistent with other SGS materials and arises from intrinsic Berry phase curvature.\cite{Jamer1,Mn2CoAlXu,FengTi,Bainsla1, Bainsla2,Mn2CoAl1} The as-grown Cr$_2$CoGa exhibits SGS behavior indicated by the small AHE coefficient and the small activation energy characteristic of a low energy band gap.\cite{Mn2CoAlXu, Bainsla1, Bainsla2}

The magnetic properties of MBE-grown Cr$_2$CoGa thin films were investigated as a function of annealing. RHEED monitoring (see Fig. 2 left inset) indicated that the Cr$_2$CoGa grew partially oriented relative to the GaAs substrate. The structural properties of the lattice showed a slight tetragonalization of the lattice parameters in the as-grown sample. When the sample was annealed above 300 $^o$C, peaks that were unrelated to the Heusler structure appeared in the diffractogram indicating phase segregation. Magnetic measurements indicated that the as-grown lattice has a low total magnetic moment (0.5 $\mu_B$/f.u.), however, the magnetic moment varied as the phase becomes segregated. The XAS spectra further supported the phase segregation of the lattice, since the valence states of the Cr changes visibly as a function of annealing. However, the as-grown samples did exhibit the Heusler structure and revealed a low magnetic moment with antiferromagnetically coupled Cr and Co atoms. The as-grown sample exhibited SGS electrical properties that could be utilized in future devices.
\squeezeup
\squeezeup
\squeezeup
 
\begin{acknowledgements}
\squeezeup
We thank T. Hussey for assistance with magnetometry and D. Arena at NSLS beamline U4B for his guidance. We thank Brian Lejuene for his help with making the contacts for electrical transport measurements. This work was primarily supported by the National Science Foundation grant ECCS-1402738. Use of the National Synchrotron Light Source, Brookhaven National Laboratory, was supported by the U.S. Department of Energy, Office of Science, Office of Basic Energy Sciences, under Contract no. DE-AC02-98CH10886.
\end{acknowledgements}
\label{References}
\bibliographystyle{unsrtnat} 
\bibliography{cr2cogabibnotitle}
\end{document}